# Which factor dominates the industry evolution? A synergy analysis based on China's ICT industry

## Yaya Li [1], Yongli Li [2], Yulin Zhao [1], Fang Wang [1]

[1] *Wuhan University of Technology*
*Gongda Rd. 25, 430070, Wuhan, P.R.China*
*e-mail: yizhi19881107@126.com; zhaoyulin_whut@126.com; wangfang_whut@126.com*

[2] *Harbin Institute of technology*
*West st. 92, 150001, Harbin, P.R.China*
*e-mail: liyongli_hit@126.com*

*Abstract*
*Industry evolution caused by various reasons, among which technology progress driving industry development has been approved, but with the new trend of industry convergence, inter-industry convergence also plays an increasing important role. Contrasting to the previous studies, this paper plans to probe the industry synergetic evolution mechanism based on industry convergence and technology progress.*
*Firstly, we use self-organization method and Haken Model to establish synergetic evolution equations, select technology progress and industry convergence as the key variables of industry evolution system; then use patent licensing data of china's listed ICT companies to measure industry convergence rate and apply DEA Malmquist index method to calculate technology progress level; furthermore apply simultaneous equation estimation method to investigate the synergetic industry evolution process. From 2002 to 2012, China's ICT industry develops rapidly; it has the most obvious convergence and powerful technology progress compared with other industries. We choose china's listed ICT industry to make empirical analysis.*
*The first main contribution of this paper lies in that our study is the first research to link technology progress and industry convergence in analyzing industry evolution and the second major contribution is to propose a novel analysis method for uncovering the mechanism of industry evolution.*
*Our main findings are: a) technology progress is the order parameter which dominates industry system evolution. Moreover, industry convergence is the control parameter which is influenced by technology progress; b) Development of technology progress is the core factor for causing evolution of industry system, and industry convergence is the outcome of technology progress; c) Especially, it is important that the dominated role of technology progress will be sustained, even though in the environment of convergence, companies also need focus on self-innovation, rather than only adapt to the new industry evolution trend.*

## Introduction

With the new industry emerging, it has undergone evolution with rapid or even radical changes. These changes often attribute to one main reason, namely technology progress. However, especially in the last two decades, industry convergence as a new and decisive phenomenon has also been found (Rosenberg, 1963) and gradually spreads throughout the whole economic studies (Dosi, 1982; Dowling, 1998; Lei, 2000; Fai & Tunzelmann, 2001). The above two aspects, technology progress and industry convergence, are often seen as the two fundamental factors for the industry growth recently (Bonnet &Yip, 2009).

As to technology progress, on the one hand, many researchers have claimed that technology progress is the driving force behind economic or industry growth. For example, in the work of Schumpeter (1934), he created the innovation theory which claimed that technical innovation promotes economic development. Solow (1956) measured technical progress contribution rate and created a new growth theory, and the similar work also can be found in Romer (1986), Lucas (1988), Grossman & Helpman (1991). Therefore, as the above literatures show, technical progress has been approved and accepted as the major sources of radical innovation within the industry, which drive the economic growth and the development of industries.

As to industry convergence, on the other hand, the definitions can go back to the early 1960s. Rosenberg (1963), based on his study on US machine tool industry, indicated that different industries relied increasingly on the same set of technological skills in their production process, and termed the set of technological skills as technological convergence. There have been many researchers studying the phenomena from different perspectives. For example, some studied different stages of convergence from the angle of evolutionary economics (Hacklin, 2010; Curran, 2011), some tried to measure convergence of different industries (Wan et al., 2011; Karvonen et al., 2012), and some investigated the convergence phenomenon with a focus on the information and communication technology (ICT) industry (Dusteers & Hagedoorn, 1988; Stieglitz, 2003; Wan et al., 2011). As Gambardella & Torrisi (1998) has claimed, during 1990s electronics sectors have undergone a trend of obvious convergence, and accordingly there is a positive correlation between technology convergence and improved performance. Recently, there also have been studies on the convergence appearing in other industries (Bröring et al., 2006; Curran et al., 2010; Karvonen & Kassi, 2012). To sum up, industry convergence becomes another global fundamental mode affecting industries and companies.

The above two theories offer logical explanation of industry evolution, but in our viewpoint, they do not cover the mechanism inside the evolution, especially ,with the trend of industry convergence, industry evolution is not only derived by technology progress within one sole industry, and on the contrary inter-industry convergence also plays an increasingly important role. Interestingly, comparing industry convergence with technology progress, which one plays a dominated role in industry evolution? And, does synergetic effect between them exist? The purpose of this research is mainly based on the two questions.

To answer the above two questions, this paper established a series of synergetic equations based on the theory of self-organization and Haken model (Haken, 1988). Whole industry is considered as a system, selected technology progress and industry convergence as two endogenous factors changing with the evolution of the system. On the one hand, there are two kinds of parameters in the Haken model: the order parameter and the control parameter. The order parameter is used to govern the evolution of system and the control parameter is dominated by the order parameter. Through this model we can distinguish which variable is the order parameter and which one is the control one; we can know which one plays the critical role in the system evolution during the concerning period. On the other hand, we can explore how the two variables affect each other. This method can appropriately address two problem mentioned above.

On the basis of the proposed equations and the Haken model, this paper further made an empirical study by using the data from China's information and communication technology (ICT) industry. The ICT industry in China has developed rapidly since 1990s, and especially enjoyed the faster growth from 2002 to 2012. Thus, the data during this period is very suitable for the empirical analysis, when we consider that the data reflects the most obvious convergence within industries and the powerful technology progress.

This paper collected china's patent licensing data to calculate listed 146 ICT industry technology convergence rate (TCR), simultaneously, gathered ICT listed company's data to measure technology progress level (TPL). Finally, simultaneous equation estimation method (precisely, the GMM time series method) is applied to empirically analyze the synergetic industry evolution process based on the established model and measured data.

The rest of the paper is structured as follows: Section 2 presents synergetic evolution model of industry system based on industry convergence and technology progress; Section 3 explains the data and the measurement of industry convergence and technology progress; Section 4 discusses the empirical results, and section 5 concludes the paper.

## Model and synergetic equations

Considering the whole ICT industry as a system, according to what we have introduced, technology progress and industry convergence are the two fundamental factors in the system. The two variables are changing synergistically with the evolution of the system. Based on the theory of evolution system, Haken model is a proper model to explore the synergetic effect of the two critical variables. The Haken model is quite famous in the field of system science but not often be used in the field of Engineering Economics, especially in the field of Industry Economics. Since the economics system is really a system in the real life, we plan to adopt the Haken model to analyze such a system.

Given two variables in an evolution system, the relationship between the two variables can be expressed by the following synergetic equations based on the Haken model:

$$dq_1/dt = -\lambda_1 q_1 - a q_1 q_2, \qquad (1)$$
$$dq_2/dt = -\lambda_2 q_2 + b q_1^2. \qquad (2)$$

Where, $q_1$ and $q_2$ are the two key variables, $\lambda_1$, $\lambda_2$, $a$ and $b$ are four parameters. According to the property of Haken model (Haken, 1998), we have the definitions of order variable and control variable as follows.

[**Definition 1**] (Order Variable and Control Variable). The variable $q_1$ is called the order variable and the variable $q_2$ is called the control variable when $\lambda_2 > 0$ and $\lambda_2 \succ |\lambda_1|$.

Often, in a real world application, the discretization forms of equation (1) and (2) are often used, since the survey data is often discrete in the unit of year, semi-year, quarter, or month, etc. Their discretization forms are

$$q_1(t+1) = (1-\lambda_1)q_1(t) - a q_1(t) q_2(t), \qquad (3)$$
$$q_2(t+1) = (1-\lambda_2)q_2(t) + b q_1(t)^2. \qquad (4)$$

Where, $t$ expresses the time period and takes the values 1, 2, 3, ….

On the basis of the discretization forms, the four parameters $\lambda_1$, $\lambda_2$, $a$ and $b$ can be calibrated based on the estimation of simultaneous equations from the theory of Econometrics. To make it strict, we rewrite the equations (3) and (4) in the form for the econometrics analysis by adding two residual terms $\varepsilon_1(t)$ and $\varepsilon_2(t)$.

$$q_1(t+1) = (1-\lambda_1)q_1(t) - a q_1(t) q_2(t) + \varepsilon_1(t), \qquad (5)$$
$$q_2(t+1) = (1-\lambda_2)q_2(t) + b q_1(t)^2 + \varepsilon_2(t). \qquad (6)$$

Then, by examining whether $\lambda_2$ is above zero and comparing $\lambda_2$ with $|\lambda_1|$, we can judge which variable is the order variable and which one is the order variable. It is noted that the method is just the way to answer the first question proposed in the Introduction part. To sum up the above analysis, we can have the following steps to determine the order variable and the control variable.

*Step 1*. Obtain the data of $q_1(t)$ and $q_2(t)$ by calculating based on the raw data from yearbook or other sources.

*Step 2.* Give the hull hypothesis: $\lambda_2 > 0$ and $\lambda_2 \succ |\lambda_1|$, which means that the $q_1$ is order variable and the $q_2$ is control variable.

*Step 3.* Estimate the equations (5) and (6) by using the approach of simultaneous equations estimation, and then calibrate the four parameters $\lambda_1$, $\lambda_2$, $a$ and $b$.

*Step 4.* Validate the hull hypothesis based on the results of the step 3. If it is accepted, we infer that the $q_1$ is order variable and the $q_2$ is control variable; whereas, if it is rejected, we should change the place of $q_1$ and $q_2$ in the equations (3) and (4), then repeat the step 3.

For better understanding the above synergetic equations (1) and (2) (noted the equations (3) and (4) are similar) and also for answering the second question, we draw the sketch map of such equations.

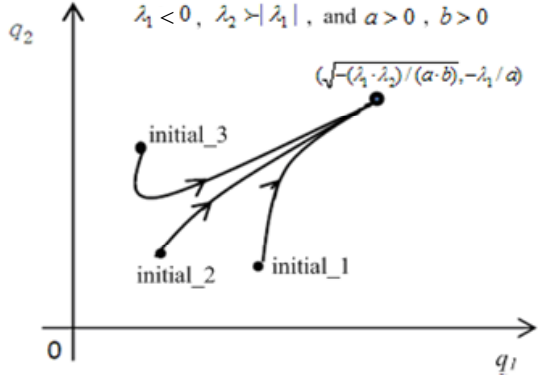

**Figure 1.** Sketch map in the first case

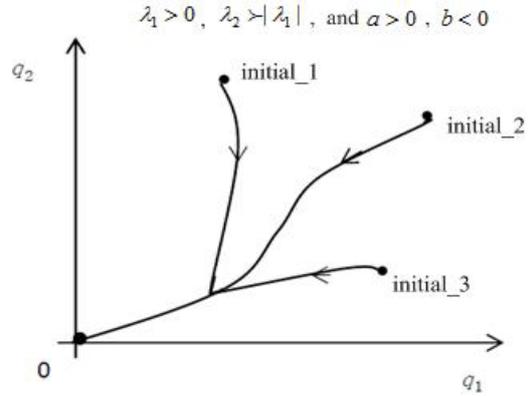

**Figure 2.** Sketch map in the second case

When satisfying that $q_1$ is order variable and the $q_2$ is control variable, namely $\lambda_2 > 0$ and $\lambda_2 \succ |\lambda_1|$, the synergetic effect between $q_1$ and $q_2$ can be very different according to different parameter values. As the Figure1 shows, the three different initial points converge into the same point A (its coordinate $(\sqrt{-(\lambda_1 \cdot \lambda_2)/(a \cdot b)}, -\lambda_1/a)$) with the time goes by. However, differences also exist among the three paths owing to different initial points. The $q_2$ beginning at the initial_3 point decreases at first and then goes up to the convergence point, whereas, in the other two paths, $q_1$ and $q_2$ are both increasing at the same time. While, because the parameters take different values, Figure2 shows the contrast trend compared to Figure1. All the three paths in Figure2 converges to the zero point (denoted by O in the figure), although the trends of them seem to be a little different similar to the situation shown in Figure 1. From the above analysis and the two sketch maps shown in Figure 1 and Figure 2, two properties can be summarized about the synergetic effect between $q_1$ and $q_2$ with different parameter values.

[**Property 1**] Given that $q_1$ is the order variable and $q_2$ is the control variable, if $\lambda_1 < 0$, $a > 0$ and $b > 0$, then point $(q_1, q_2)$ will converge to $(\sqrt{-(\lambda_1 \cdot \lambda_2)/(a \cdot b)}, -\lambda_1/a)$ with time goes by, although their paths can show different trends at the beginning according to the different initial points.

[**Property 2**] Given that $q_1$ is the order variable and $q_2$ is the control variable, if $\lambda_1 > 0$, $a > 0$ and $b < 0$, then the point $(q_1, q_2)$ will converge to $(0,0)$ with time going by, although their paths can show different trends at the beginning according to the different initial points.

We need to explain that the two converge points $(\sqrt{-(\lambda_1 \cdot \lambda_2)/(a \cdot b)}, -\lambda_1/a)$ and $(0,0)$ can be calculated out from the following steady state equations (7) and (8) induced from the equations (1) and (2), respectively.

$$-\lambda_1 q_1 - a q_1 q_2 = 0, \quad (7)$$
$$-\lambda_2 q_2 + b q_1^2 = 0. \quad (8)$$

It is noted that the cases of different parameter values are not confined to the above two kinds, for example, $\lambda_1 > 0$, $a < 0$ and $b > 0$, but the other cases are not very useful in our analysis, especially considering the actual data prepared for the forthcoming empirical analysis. This will be showed further in the part of empirical analysis.

After the equations (3) and (4) has been calibrated based on the estimation of simultaneous equations, we can obtain the estimation values of the four parameters and judge which situation (or called which figure) the synergetic effect between $q_1$ and $q_2$ belongs to. From the sketch map shown in the corresponding figure, we can get the answer of the second question that how the two variables affect each other, namely the synergetic effect.

**Data preparation**

Technology progress (TP) and industry convergence (IC) are needed to be measured firstly. They are the necessary data for further analysis and also reflect the development levels of technology progress and industry convergence in recent years. In the Introduction of this paper, we have explained why we select China's data for this problem. In this section, we will show the details of how to measure these two variables. In all, it is an elaborated process needing more painstaking work.

Technology progress (TP) is measured by using DEA-based Malmquist index method, whose result is called the technology progress level (TPL). Such method was proposed by Fare et al (1994) and was found based on the DEA approach (see also Sufian, et al, 2010; Li, et al (2013)). The Malmquist index is defined on the basis of distance functions presented by Caves, et al (1982), and it measures the change of the total factor productivity (TFP) between two data points by calculating the ratio of their distances relative to a common technology. The total is decomposed into two parts: technical efficiency level and technological change. Here, what we concern is the technical efficiency level other than the technological change. Accordingly, the original Malmquist index method is revised for our purpose and the DEA-based Malmquist index can be obtained by utilizing the following formula:

$$\frac{TFP_{t+1}}{TFP_t} = \left( \frac{D_0^t(\mathbf{x}^{t+1}, y^{t+1})}{D_0^t(\mathbf{x}^t, y^t)} \right)^{0.5}, \quad (9)$$

where,

$$S^t = \{(\mathbf{x}^t, y^t) : \mathbf{x}^t \text{ can produce } y^t\};$$

$$D_0^t(\mathbf{x}^t, y^t) = \inf\{\theta : (\mathbf{x}^t, \frac{y^t}{\theta}) \in S^t\};$$

and

$$D_0^t(\mathbf{x}^{t+1}, y^{t+1}) = \inf\{\theta : \frac{\mathbf{x}^{t+1}}{\theta} \in S^t\}.$$

Specifically, the input $\mathbf{x}$ consists of labor and capital, the two factors, where labor is reflected by the number of employees and capital is reflected by the net fixed assets investment in the above model, and also the output $y$ is expressed by the operating income. The software DEAP 2.1 is a suitable tool for solving the above problem.

Considering the availability, the completeness and the reliability of the data, we select the listed companies of China as the analysis sample. So far, there are 146 ICT listed companies in china's two main stock markets. Table 1 shows their codes and Table 2 shows their amounts in different years.

Table 1

**Stock codes of 146 listed ICT companies in china's two main stock market**

| Stock codes | | | | | | | | | |
|---|---|---|---|---|---|---|---|---|---|
| 000021 | 000938 | 600355 | 600845 | 600476 | 002153 | 002281 | 002401 | 300038 | 300085 |
| 000035 | 000948 | 600498 | 600850 | 600570 | 002161 | 002296 | 002405 | 300042 | 002446 |
| 000063 | 000977 | 600536 | 600855 | 600990 | 002184 | 002308 | 002410 | 300044 | 002465 |
| 000066 | 000997 | 600588 | 600050 | 002052 | 002194 | 002312 | 002416 | 300045 | 002467 |
| 000070 | 600037 | 600601 | 600271 | 002063 | 002195 | 002313 | 002417 | 300047 | 300096 |
| 000503 | 600076 | 600621 | 600392 | 002065 | 002214 | 002315 | 002421 | 300050 | 300098 |
| 000547 | 600105 | 600640 | 600485 | 002089 | 002230 | 002316 | 002439 | 300051 | 300101 |
| 000555 | 600118 | 600687 | 600487 | 002090 | 002231 | 002331 | 002449 | 300052 | 300102 |
| 000561 | 600122 | 600718 | 600522 | 002093 | 002232 | 002339 | 300002 | 300059 | 300104 |
| 000586 | 600130 | 600728 | 600571 | 002095 | 002236 | 002362 | 300010 | 300065 | 300113 |
| 000701 | 600171 | 600756 | 002017 | 002104 | 002253 | 002368 | 300017 | 300074 | 300162 |
| 000748 | 600198 | 600764 | 002027 | 002106 | 002261 | 002373 | 300025 | 300075 | |
| 000823 | 600288 | 600770 | 600403 | 002115 | 002268 | 002376 | 300028 | 300076 | |
| 000851 | 600289 | 600775 | 600410 | 002148 | 002279 | 002383 | 300033 | 300079 | |
| 000892 | 600345 | 600776 | 600446 | 002151 | 002280 | 002396 | 300036 | 300081 | |

Source: selected by the authors based on the listed company information (website: http://www.cninfo.com.cn/)

Table 2

**China's listed ICT company amounts during 2002-2012**

| Year | 2002 | 2003 | 2004 | 2005 | 2006 | 2007 | 2008 | 2009 | 2010 | 2011 | 2012 |
|---|---|---|---|---|---|---|---|---|---|---|---|
| Numbers of Listed ICT Company | 48 | 55 | 63 | 63 | 64 | 73 | 86 | 88 | 136 | 146 | 146 |

Source: selected by the authors based on the listed company information (website: http://www.cninfo.com.cn/)

All the data (mainly $(\mathbf{x}^t, y^t)$ of each year) is collected from 146 listed ICT companies' semi-annual reports from 2002 to 2012. Two things have been paid much attention to in our calculation process. The first one is that the data in the first half year of 2002 is taken as the benchmark, and all the years' TPLs are further calculated based on this year. The purpose of doing this is to make all the TPLs have the same benchmark so that they are comparable at different years. The approach of calculating is shown in formula (10), where $n = 0.5, 1, 1.5, 2, \cdots$. And, the second one is that not every listed company exists at the beginning of the survey period, namely some of them went public later than some others. Thus, we make the data start year as the measurement-based year, calculate every listed ICT company's TPL in their public duration, and then get the average of all the ICT companies' TPLs existing in the corresponding semi-year to be the result. All the results are listed in Table 3. As the table illustrates, China's listed ICT industry undergoes obviously improving technology progress from 2002 to 2012.

$$\text{TPL}_{2002+n} = \frac{TFP_{2002+n}}{TFP_{2002}} = \frac{TFP_{2002+0.5}}{TFP_{2002}} \cdot \frac{TFP_{2002+1}}{TFP_{2002+0.5}} \cdots \frac{TFP_{2002+n}}{TFP_{2002+n-0.5}}. \qquad (10)$$

Table 3

**China's listed ICT industry TPL from 2002 to 2012**

| Year | 2002 fh | 2002 sh | 2003 fh | 2003 sh | 2004 fh | 2004 sh | 2005 fh | 2005 sh | 2006 fh | 2006 sh | 2007 fh |
|---|---|---|---|---|---|---|---|---|---|---|---|
| TPL | 1.000 | 1.097 | 1.163 | 1.161 | 1.226 | 1.248 | 1.253 | 1.273 | 1.266 | 1.268 | 1.274 |
| Year | 2007 sh | 2008 fh | 2008 sh | 2009 fh | 2009 sh | 2010 fh | 2010 sh | 2011 fh | 2011 sh | 2012 fh | 2012 sh |
| TPL | 1.301 | 1.269 | 1.303 | 1.324 | 1.369 | 1.378 | 1.391 | 1.401 | 1.435 | 1.456 | 1.523 |

Source: calculated by the authors based on the company semi-year reports
Notes: fh denotes the first half year; sh denoted the second half year.

We next present how to measure the industry convergence (IC). Generally speaking, there are two streams of literatures on measuring industry convergence. One stream measures industry diversification (Teece, et al, 1994; Gambardella & Torrisi, 1988), and the other measures technology relatedness (Fai & Tunzelmann, 2001). Our paper follows the latter stream, which uses patent data to analyze convergence. The technology relatedness is always taken as a significant measure for industry technology convergence (Karvonen et al, 2012), and furthermore the patent licensing data is suitable for illustrating inter-industry technology convergence (Geum, Kim, Lee & Kim, 2012), which is proper for our problem.

From China's state intellectual property office (SIPO), we have got the material called *patent licensing contract records information table* which records the basic information of one patent including Patent Number, Invention Name, Grantor, Grantee and Record Time. One example has been given in Figure 3. We selected data of the listed ICT companies, which have been mentioned above, from the *information table*, and the whole data period is from 2002 to 2012.

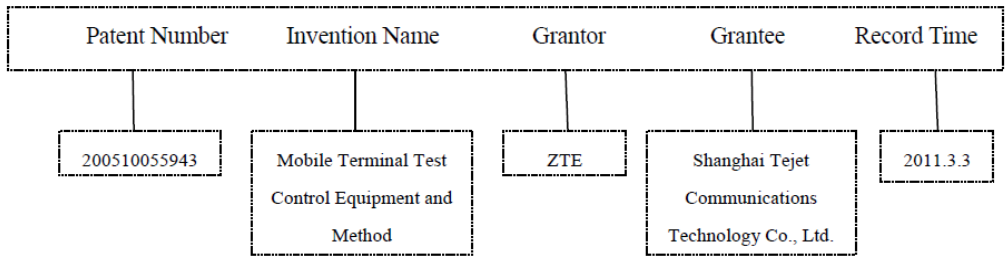

**Figure 3** One example of the patent record in the *patent licensing contract records information table*

All the listed ICT companies can be divided into seven sectors: the communication equipment manufacturing sector (CEM), the computer manufacturing sector (CM), the electronic components manufacturing sector (ECM), the household and video equipment manufacturing sector (HEM), the other electronic equipment manufacturing sector (OM), the information transmission service sector (IS), and the computer service and software sector (CS). The technology relatedness can be measured by using one transfer matrix as shown in Table 4. Here, we take the data in 2012fh (the first half year) as an example. It is noted that we make half a year as an investigation unit.

Table 4

**ICT industry technology transfer matrix of china in 2012fh**

| ICT sectors | CEM | CM | ECM | HEM | OM | IS | CS |
|---|---|---|---|---|---|---|---|
| CEM | 7 | 2 | 1 | 1 | 1 | 1 | 2 |
| CM | 2 | 1 | 2 | 0 | 1 | 0 | 1 |
| ECM | 1 | 2 | 13 | 2 | 0 | 2 | 0 |
| HEM | 0 | 1 | 2 | 15 | 3 | 0 | 2 |
| OM | 1 | 0 | 1 | 3 | 4 | 3 | 1 |
| IS | 0 | 1 | 0 | 0 | 0 | 2 | 0 |
| CS | 1 | 0 | 1 | 2 | 2 | 1 | 7 |

Source: elaborated by the authors based on th*e patent licensing contract records information table* of china in 2012

The table has 7 rows and 7 columns. Each row represents one sector's patent penetrating into the other ICT sub-sectors, and each column illustrates patents absorbed from the responding ICT sub-sectors. The data on the diagonal line of matrix reflects technology flow within the same sector, and the data not on the diagonal line expresses technology convergence between different sectors. From the Table 4, we can find that 7 patents flowed within CEM sector in 2012, while 2 patents from CEM sector are absorbed by CM sector. Based on the above matrix, industry technology convergence ratio (TCR) can be calculated by the following formula:

$$\text{TCR}_j^n = \left(\sum_{i=1}^{7} a_{ij}^n - a_{jj}^n\right) \bigg/ \sum_{i=1}^{7}\sum_{j=1}^{7} a_{ij}^n, \qquad (11)$$

where, $a_{ij}^n$ donates the number of the row *i* and column *j* in the matrix of year *n*, and the parameter *j* takes values from 1 to 7, which reflects the seven sectors from CEM to CS in sequence. Then, the whole year's $TCR^n$ is the sum of all the individual sectors' $TCR_j^n$. As a result, we obtain the $TCR^n$ year by year as Table 5 shows.

Table 5

**China's ICT industry technology convergence ratio (TCR) from 2002 to 2012**

| Year | 2002 fh | 2002 sh | 2003 fh | 2003 sh | 2004 fh | 2004 sh | 2005 fh | 2005 sh | 2006 fh | 2006 sh | 2007 fh |
|---|---|---|---|---|---|---|---|---|---|---|---|
| TCR | 0.1091 | 0.1386 | 0.1667 | 0.1759 | 0.2190 | 0.2330 | 0.2098 | 0.2583 | 0.2798 | 0.2857 | 0.2902 |
| Year | 2007 sh | 2008 fh | 2008 sh | 2009 fh | 2009 sh | 2010 fh | 2010 sh | 2011 fh | 2011 sh | 2012 fh | 2012 sh |
| TCR | 0.3622 | 0.3455 | 0.3668 | 0.3783 | 0.3788 | 0.3945 | 0.4059 | 0.4076 | 0.4130 | 0.5158 | 0.4704 |

Source: calculated by the authors based on the method proposed.

**Empirical Results**

Based on Data Preparation, the empirical analysis can be carried out to answer the two questions proposed in the Introduction. Before estimation methods of simultaneous equations are selected and discussed, the two-dimensional diagram consisting of China's ICT industry technology convergence rate (TCR) and technology progress level (TPL) is drawn to show the visual trend during the period from 2002 to 2012.

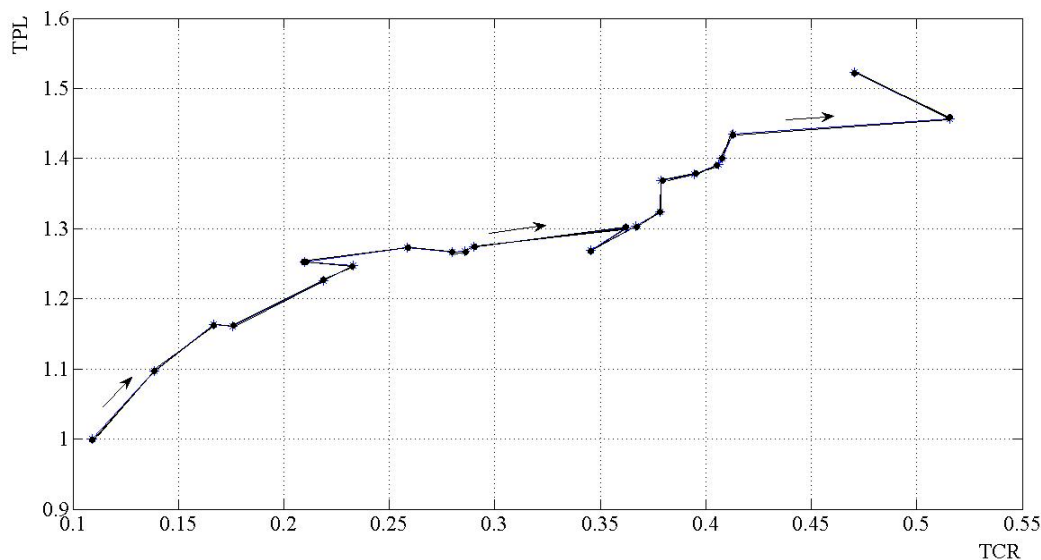

**Figure 4** The two-dimensional diagram of TCR and TPL

In the Figure 4, the TCR is horizontal axis and the TPL is the vertical axis, and the directions of arrows express the evolutions of the two series of variables with the years going by. On the whole, the two variables show a basically increasing trend, although some fluctuations exist. However, the further technique tool is necessary to answer the two questions, since the above graph can not tell which variable the order one is and which the control one is.

As the equations (5) and (6) show, the method for estimating the simultaneous equations is needed. Usually, three approaches are used to the system estimation of simultaneous equations, and they are three-stage least squares (3SLS) (Zellner et al, 1962), full information maximum likelihood (FIML) (Enders, 2001), and generalized method of moments (GMM) (Stock et al, 2002; Newey et al, 2009), respectively. We need to choose the most proper one from them to estimate the equations (5) and (6). It is obvious that the data is the kind of time series data, thus GMM method is the most suitable one because GMM allows the random error items of simultaneous equations have heteroscedasticity and serial correlation, while the other two methods do not allow this. Besides, GMM has the lower restriction about the distribution of the random errors compared to the other two methods and it is also a robust estimation method. Thus, facing the features of the data selected in our study, we regard the GMM as a good method to estimate the equations.

Following the steps, hull hypothesis should be given first. Firstly, we assume TCR is the order variable and TPL is the control variable, namely TCR is taken as the $q_1(t)$ and TPL as the $q_2(t)$ in the equations (5) and (6). Then, it is needed to examine whether $\lambda_2 > 0$ and $\lambda_2 \succ |\lambda_1|$ according to Definition 1 given in this paper. Here, in the process of estimation, $q_1(t-1)$, $q_2(t-1)$ ($t = 2, 3, \cdots$) and the constant term are considered as the tool variables, whereas $q_1(t)$ and $q_2(t)$ are considered as the endogenous variable. Because the number of tool variables is larger than the parameters, the whole system of equations is identifiable. Table 6 gives the results of the above assumption via Eviews 5.1 (using the GMM-time series (HAC) function) as below.

Table 6

**Estimation results (TCR assumed to be order variable, TPL to be control variable)**

| parameters | coefficient | std. error | t-statistic | prob. | Equation $R^2$ |
|---|---|---|---|---|---|
| $\lambda_1$ | -0.677940 | 0.166835 | -4.063531 | 0.0002 | 0.900249 |
| $a$ | 0.474084 | 0.131757 | 3.598157 | 0.0009 | |
| $\lambda_2$ | -0.027792 | 0.010476 | -2.652855 | 0.0116 | 0.908468 |
| $b$ | -0.110465 | 0.116303 | -0.949796 | 0.3482 | |

From Table 6, the coefficients of $\lambda_1$ and $\lambda_2$ do not satisfy the conditions that $\lambda_2 > 0$ and $\lambda_2 \succ |\lambda_1|$. Thus, the null hypothesis is rejected, that is to say, the estimation results can not support that TCR is the order variable and TPL is the control variable. However, the above result can not prove that its converse is right, because, in the system consisting of equations (5) and (6), it is likely that no variable is the order variable. Thus, we need to further examine the other null hypothesis that TPL is the order variable and TCR is the control variable. At this time, TPL is taken as the $q_1(t)$ and TCR as the $q_2(t)$ in the equations (5) and (6). The estimation results by the GMM-time series are listed in Table 7 and the fitted graph is shown in Figure 5 with the real numbers of TCR and TPL during the 11 years.

Table 7

**Estimation results (TPL assumed to be order variable, TCR to be control variable)**

| parameters | coefficient | std. error | t-statistic | prob. | Equation $R^2$ |
|---|---|---|---|---|---|
| $\lambda_1$ | -0.036735 | 0.017463 | -2.103608 | 0.0421 | 0.913313 |
| $a$ | 0.056938 | 0.051895 | 1.097183 | 0.2795 | |
| $\lambda_2$ | 0.207174 | 0.063130 | 3.281687 | 0.0022 | 0.919055 |
| $b$ | 0.048658 | 0.011778 | 4.131242 | 0.0002 | |

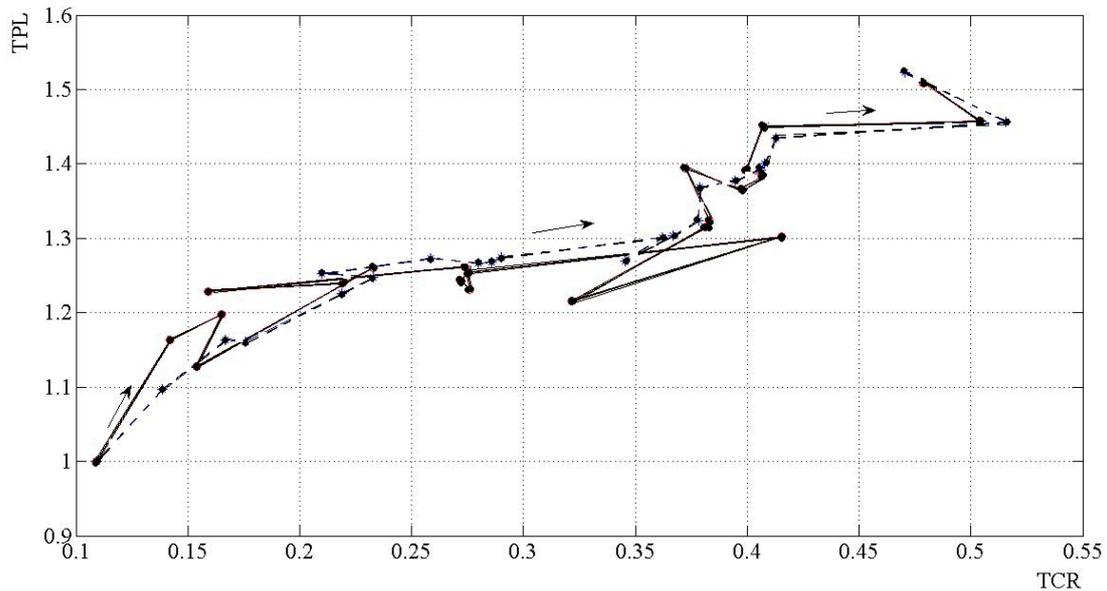

Figure 5 Real numbers and fitted numbers of the two variables

At this time, $\lambda_2 > 0$ and $\lambda_2 \succ |\lambda_1|$ holds as shown in Table 7, which means TPL can be seen as the order variable and TCR as the control variable according to the estimation results. Besides, judged from the values of the t-statistic and the prob. index, the coefficients of three parameters $\lambda_1$, $\lambda_2$ and $b$ are significant at the confidence level of 0.05. Also, the two equations both have a high $R^2$ which are more than 0.90. The Figure 5 shows the fitted numbers and the real numbers, from which we can find the two lines are close to each other. In the whole, the estimation results are satisfied.

Based on the estimation results, the equations (5) and (6) can be calibrated as

$$\begin{cases} \text{TPL}(t+1) = 1.036735 \cdot \text{TPL}(t) - \underline{0.056938} \cdot \text{TPL}(t) \cdot \text{TCR}(t) \\ \text{TCR}(t+1) = 0.792826 \cdot \text{TCR}(t) + 0.048658 \cdot \text{TPL}(t)^2 \end{cases},$$

Where, the underlined coefficient is not significant in the statistical sense. To sum up the results and the above simultaneous equation, two answers can be given in allusion to the two questions of this paper:

1) TPL is the order variable and TCR is the control variable; accordingly, the whole system is mainly affected by the TPL. In other words, the development of technical progress (reflected by TPL) is the core factor for causing the evolution of the whole system.

2) TCR is dominated by TPL according to the second equation, whereas TCR does not affect TPL significantly since the underlined coefficient in the first equation is not significant. The relationship between the two variables indicate their synergetic effects, namely the industry technical convergence (reflected by TCR) is the outcome of the technical progress (reflected by TPL), but the industry technical convergence in itself does not take obvious effect on the technical progress during the period we have used for empirical analysis.

**Conclusions**

The main objective of this study was to answer the two questions that (1) comparing industry convergence with technology progress, which one plays a dominated role in industry evolution, and (2) how industry convergence and technology progress synergistically affect industry evolution. The first contribution of this paper lies in the fact that our study is the first research to link technology progress and industry convergence in analyzing industry evolution. In this explorative study, we have examined industry evolution mechanism based on industry convergence and technology progress.

The second major contribution of this paper is to propose a novel analysis method for uncovering the mechanism of industry evolution. We applied self-organization method and Haken Model to establish synergetic evolution equations, and selected industry convergence and technology progress as the key variables of industry evolution system. Furthermore, we collected china's patent licensing data to calculate listed ICT industry technology convergence rate, simultaneously, we gathered 146 ICT listed company's data to measure technology progress level. Finally, simultaneous equation estimation method (GMM time series method) is applied to empirically analyze the synergetic industry evolution process based on the established model and measured data. The conclusions and policy implications are as follows.

From 2002 to 2012, China's ICT industry enjoys increasing convergence rate and improving technology progress level. As for the whole industry system, technology progress and industry convergence are the two endogenous factors affecting the system evolution. Our result indicates that the development of technology progress is the core factor for causing evolution of industry system, and industry convergence is the outcome of technology progress. In other words, industry synergetic evolution mechanism can be summarized that (1) technology progress is the order parameter which dominates the evolution of system, and (2) industry convergence is the control parameter which is reflected by technology progress.

The policy implication of our result is important for companies and government. It is important that the dominated role of technology progress will be sustained, even though in the environment of convergence, companies also need focus on self-innovation, rather than only adapt to the new industry evolution trend.

Although the results are achieved from China's ICT industry data, they may have implications for other countries. Limited by data availability, this paper only investigated the listed ICT industries from 2002 to 2012. Exploratory work on other industries and international comparisons would be directions for future research.

**References**

Bröring, S. (2005). The Front End of Innovation in Converging Industries: The Case of Nutraceuticals and Functional Foods. *Wiesbaden, Germany*: DUV.

Caves, D.W., Christensen, L.R., & Diewert, W. E. (1982). The Economic Theory of Index Numbers and the Measurement of Input, Output and Productivity. *Econometrica,* 50(6), 1393-1414.

Curran, C. S., Bröring, S., & Leker, J. (2010). Anticipating Converging Industries Using Publicly Available Data. Technological Forecasting *&* Social Change, 77(3), 385 -395. http://dx.doi.org/10.1016/j.techfore.2009.10.002

Curran, C. S., & Leker, J. (2011). Patent Indicators for Monitoring Convergence - Examples from NFF and ICT. *Technological Forecasting & Social Change*,78(2) , 256 -273. http://dx.doi.org/10.1016/j.techfore.2010.06.021

Dosi, G. (1982). Technological Paradigms and Technological Trajectories: A Suggested Interpretation of the Determinants and Directions of Technology Change. *Research Policy*, 11(3), 147-163. http://dx.doi.org/10.1016/0048-7333(82)90016-6

Dowling, M., Lechner, C., & Thielman, B. (1998). Convergence: Innovation and Change of Market Structures between Television and Online services. *Electronic Markets,* 8(4), 31-35. http://dx.doi.org/10.1080/10196789800000053

Duysters, G., & Hagedoorn, J. (1998).Technological Convergence in the IT Industry: The Role of Strategic Technology Alliances and Technological Competencies. *International Journal of Economics and Business,* 5(3), 355-368. http://dx.doi.org/10.1080/13571519884431

Enders, C. K. (2001). The Performance of the Full Information Maximum Likelihood Estimator in Multiple Regression Models with Missing Data. *Educational and Psychological Measurement,* 61(5), 713-740. http://dx.doi.org/10.1177/0013164401615001

Fai, F. M., & Tunzelmann, N.V. (2001). Industry-Specific Competencies and Converging Technological Systems: Evidence from Patents. *Structural Change and Economic Dynamics,* 12(2), 141-170.
http://dx.doi.org/10.1016/S0954-349X(00)00035-7

Färe, R., Grosskopf, S., Norris, M. & Zhang, Z. (1994). Productivity Growth, Technical Progress and Efficiency Changes in Industrialised Countries. *American Economic Review,* 84, 66-83.

Gambardella, A., & Torrisi, S. (1998). Does Technological Convergence Imply Convergence in Markets? Evidence from the Electronics Industry. *Research Policy,* 27(5), 445-463. http://dx.doi.org/10.1016/S0048-7333(98)00062-6

Geum, Y., Kim, C., Lee, S. & Kim, M. (2012).Technological Convergence of IT and BT: Evidence from Patent Analysis. *ETRI Journal,* 34(3), 439 -449.http://dx.doi.org/10.4218/etrij.12.1711.0010

Grossman, G. M., & Helpman, E. (1991). Quality Ladders in the Theory of Growth. *Review of Economic Studies,* 58(1), 43-61. http://dx.doi.org/10.2307/2298044

Haken, H. (1998). Information and Self-Organization: A Macroscopic Approach to Complex System. *Berlin & New York :Springer-verlag,* 1988, 134-167.

Hacklin, F., Marxt, C. & Fahrni, F. (2010). An Evolutionary Perspective on Convergence: Inducing a Stage Model of Inter - industryInnovation. *International Journal of Technology Management,* 49, 220-249.
http://dx.doi.org/10.1504/IJTM.2010.029419

karvonen, M., Lehtowaara, M., & Kassi, T. (2012). Build-up of Understanding of Technological Convergence: Evidence from Printed Intelligence Industry. *International Journal of Innovation and Technology Management,* 9(3) ,1094-1107.
http://dx.doi.org/10.1142/S0219877012500204

karvonen, M., & kassi, T. (2010). Analysis of Convergence in Paper and Printing Industry. *Journal of Engineering Management and Economics,* 1(4), 269-293.
http://dx.doi.org/10.1504/IJEME.2010.038647

Kim, M. S., & Kim, C. (2012). On a Patent Analysis Method for Technological Convergence. *Procedia social and Behavioral sciences,* 40, 657-663. http://dx.doi.org/10.1016/j.sbspro.2012.03.245

Lei, D. T. (1956). Industry Evolution and Competence Development:The Imperatives of Technological Convergence. *International Journal of Technology Management*, 19(78), 699-738.

Lucas, R. (1988). On the Mechanics of Economic Development. *Journal of Monetary Economics*, 22, 342.
http://dx.doi.org/10.1016/0304-3932(88)90168-7

Newey, W. K., & Windmeijer, F. (2009). GMM with Many Weak Moment Conditions. *Econometrica,* 77, 687-719.
http://dx.doi.org/10.3982/ECTA6224

Pennings, J.M., & Puranam, P. (2001). Market Convergence and Firm Strategy: New Directions for Theory and Research. In paper presented at the ECIS conference, Eindhoven, Netherlands.

Rosenberg, N. (1963).Technological Change in the Machine-Tool Industry,1840-1910. *The Journal of Economic History,* 23(4), 414-443.

Romer, P. M. (1986). Increasing Returns and Long-Run Growth. *Journal of Political Economy,* 94(5), 1002- 1037.

Stieglitz, N. (2003). Digital Dynamics and Types of Industry Convergence: the Evolution of the Handheld Computer Market. In J. F. Christensen (Ed.), The industrial dynamics of the new digital economy (pp. 179-208). Northampoton, MA: Edward Elgar Publishing.

Stock, J. H., Wright, J. H., & Yogo, M. (2002). A Survey of Weak Instruments and Weak Identification in Generalized Method of Moments. *Journal of Business and Economics Statistics,* 20, 518-529.
http://dx.doi.org/10.1198/073500102288618658

Schumpeter, J. (1934).The Theory of Economic Development.Cambridge:Harvard University Press

Solow, R. M. (1956). A Contribution to the Theory of Economic Growth. *Quarterly Journal of Economics,* 70(1), 65-94.
http://dx.doi.org/10.2307/1884513

Sufian, F., & Habibullah ,M.S. (2010). Does Foreign Banks Entry Fosters Bank Efficiency? Empirical Evidence from Malaysia. *Inzinerine Ekonomika-Engineering Economics,* 21(5), 464-474.

Teece, D., Rumelt, R., Dosi ,G., & Winter ,S. (1994). Understanding corporate coherence: Theory and evidence. *Journal of Economic Behavior and Organization,* 23, 1-30.
http://dx.doi.org/10.1016/0167-2681(94)90094-9

Tushman, M. L. & Anderson, P. (1986). Technological Discontinuities and Organizational Environments. *Administrative Science Quarterly,* 31, 439-465.

Wan, X., Xuan,Y., & Lv, K. (2011). Measuring Convergence of China's ICT Industry: An Input -output Analysis. *Telecommunications Policy,* 35( 4), 301 -313. http://dx.doi.org/10.1016/j.telpol.2011.02.003

Wu, C., Li, Y. L., Liu Q., Wang, K. S. (2013). A Stochastic DEA Model Considering Undesirable Outputs with Weak Disposability. *Mathematical and Computer Modelling*, 58(5-6), 980–989.

http://dx.doi.org/10.1016/j.mcm.2012.09.022

Yip, G. S., & Bonnet, D. (2009). Strategy Convergence. *Business Strategy Review,* 20(2), 50-55.

http://dx.doi.org/10.1111/j.1467-8616.2009.00599.x

Zellner, A., & Theil, H. (1962).Three-Stage Least Squares: Simultaneous Estimation of Simultaneous Equations. *Econometrica,* 30(1), 54-78.

**Yaya Li, Yongli Li, Yulin Zhao, Fang Wang**

**Which factor dominates the industry evolution? A synergy analysis based on China's Information and Communications Technology (ICT) industry**

Summary

Industry evolution caused by various reasons, among which technology progress driving industry development has been approved (Solow, 1956; Romer 1986), but with the new trend of industry convergence (Rosenberg, 1963), inter-industry convergence also plays an increasing important role (Dosi, 1982; Dowling, 1998; Lei, 2000; Fai & Tunzelmann, 2001). The above two aspects, technology progress and industry convergence, are often seen as the two fundamental factors for the industry growth recently (Bonnet &Yip, 2009). Contrasting to the previous studies, this paper probed the industry synergetic evolution mechanism based on industry convergence and technology progress.

Even though the existing theories offer logical explanation of industry evolution, but in our viewpoint, they do not uncover the mechanism inside the evolution, especially when we mention that the growth of industry is not only derived by technology progress within one sole industry, and on the contrary inter-industry convergence also plays an increasingly important role. Comparing industry convergence with technology progress, which one plays a dominated role in industry evolution? And, does synergetic effect between them exist? The purpose of this research is mainly based on the two questions.

To answer the above two questions, this paper established a series of synergetic equations based on the theory of self-organization and Haken model (Haken, 1988). In this paper, the whole industry is considered as a system, technology progress and industry convergence can be seen as two endogenous factors changing with the evolution of the system. Actually, the theory of self-organization is a suitable method to depict the evolution system. On the one hand, there are two kinds of parameters in the Haken model: the order parameter and the control parameter. The order parameter is used to govern the evolution of system and the control parameter is dominated by the order parameter. Using this model, we can distinguish which variable is the order parameter and which one is the control one, and we can know which one plays the critical role in the system evolution during the concerning period.

On the basis of the proposed equations and the Haken model, this paper further made an empirical study by using the data from China's information and communication technology (ICT) industry. The ICT industry in China has developed rapidly since 1990s, and especially enjoyed the faster growth from 2002 to 2012, and also the data reflects the most obvious convergence within industries and the powerful technology progress.

Specifically, we applied self-organization method and Haken Model to establish synergetic evolution equations, and selected industry convergence and technology progress as the key variables of industry evolution system. Furthermore, we collected china's patent licensing data to calculate listed ICT industry technology convergence rate (TCR), and simultaneously, we gathered 146 ICT listed company's data to measure technology progress level (TPL). Finally, simultaneous equation estimation method (precisely, the GMM time series method) is applied to empirically analyze the synergetic industry evolution process based on the established model and measured data. The conclusions and policy implications are as follows.

From 2002 to 2012, China's ICT industry enjoys increasing convergence rate and improving technology progress level. As for the whole industry system, technology progress and industry convergence are the two endogenous factors affecting the system evolution. Our result indicates that the development of technology progress is the core factor for causing evolution of industry system, and industry convergence is the outcome of technology progress. In other words, industry synergetic evolution mechanism can be summarized that (1) technology progress is the order parameter which dominates the evolution of system, and (2) industry convergence is the control parameter which is reflected by technology progress.

The policy implication of our result is important for companies and government. The findings suggest that, on the one hand, companies should increase R&D input, improve self-creative ability, and enhance industry technology progress level; simultaneously, enterprises should absorb other industry innovation outcomes, and improve technology absorbing capacity. On the other hand, government authorities should adopt policies to encourage industry self-innovation, meanwhile enhance the establishment of industry generic technology platform, and improve inter-industry technology penetration and transfer. Especially, it is important that the dominated role of technology progress will be sustained, even though in the environment of convergence, companies also need focus on self-innovation, rather than only adaption to the new industry evolution trend.

The first main contribution of this paper lies in that our study is the first research to link technology progress and industry convergence in analyzing industry evolution and the second major contribution is to propose a novel analysis method for uncovering the mechanism of industry evolution. Although the results are achieved from China's ICT industry data, they may have implications for other countries. Limited by data availability, this paper only investigated the listed ICT industries from 2002 to 2012. Exploratory work on other industries and international comparisons would be directions for future research.

Keywords: Industry evolution, industry convergence, technology progress, synergy analysis, Haken Model, Information and Communications Technology industry, Patent Data, simultaneous equation estimation, China